\documentclass[a4paper,fleqn]{cas-dc}

\usepackage[authoryear,longnamesfirst]{natbib}

\usepackage{amsmath,amsfonts,bm}

\def\tsc#1{\csdef{#1}{\textsc{\lowercase{#1}}\xspace}}
\tsc{WGM}
\tsc{QE}
\tsc{EP}
\tsc{PMS}
\tsc{BEC}
\tsc{DE}

\begin{document}
\let\WriteBookmarks\relax
\def\floatpagepagefraction{1}
\def\textpagefraction{.001}
\shorttitle{Episodic memory governs choices: an RNN-based reinforcement learning model for decision-making task}
\shortauthors{Xiaohan Zhang et~al.}

\title [mode = title]{Episodic memory governs choices: an RNN-based reinforcement learning model for decision-making task}

\author[1]{Xiaohan Zhang}
\address[1]{School of Mathematics, South China University of Technology, Guangzhou, China}
\author[2]{Lu Liu}
\address[2]{Centre for Artificial Intelligence, University of Technology Sydney, Sydney, Australia}

\author[2]{Guodong Long}
\author[2]{Jing Jiang}

\author[1]{Shenquan Liu}

\ead{mashqliu@scut.edu.cn}
\cormark[1]
\cortext[cor1]{Corresponding author at School of Mathematics, South China University of Technology, Guangzhou, China}











\begin{abstract}
Typical methods to study cognitive function are to record the electrical activities of animal neurons during the training of animals performing behavioral tasks. 
A key problem is that they fail to record all the relevant neurons in the animal brain. To alleviate this problem, we develop an RNN-based Actor-Critic framework, which is trained through reinforcement learning (RL) to solve two tasks analogous to the monkeys' decision-making tasks.
The trained model is capable of reproducing some features of neural activities recorded from animal brain, or some behavior properties exhibited in animal experiments, suggesting that it can serve as a computational platform to explore other cognitive functions.   
Furthermore, we conduct behavioral experiments on our framework, trying to explore an open question in neuroscience: which episodic memory in the hippocampus should be selected to ultimately govern future decisions. We find that the retrieval of salient events sampled from episodic memories can effectively shorten deliberation time than common events in the decision-making process. The results indicate that salient events stored in the hippocampus could be prioritized to propagate reward information, and thus allow decision-makers to learn a strategy faster.

\end{abstract}


\maketitle

\section{Introduction}\medskip

A hallmark of animal brain is the capability of forming decisions from sensory inputs to guide meaningful behavioral responses. Understanding the relationship between behavioral responses and how they are encoded in brains is a major goal in the neuroscience.
To this end, behavior training of nonhuman primates has been studied in a variety of decision tasks, such as perceptual discrimination \citep{shadlen2001neural}. 
These electrophysiological experiments have uncovered that neural signals at the single-neuron level are correlated with specific aspects of decision computation. However, in the mammalian brain, a decision is made not by a single neuron, but by the collective dynamics of neural circuits. Unfortunately, the animal-based experiment does not allow us to access all of the relevant neural circuits in the brain. To address this problem, neural circuit modeling with recurrent neural network has been used to uncover circuit mechanisms underlying complex behaviors \citep{mante2013context}. 

The contributions of the prefrontal cortex-basal ganglia to complex behaviors are still not completely understood. A wide array of evidence~\citep{o2004dissociable,sohal2009parvalbumin} shows that the prefrontal cortex-basal ganglia circuit appears to implement RL algorithm and is driven by a reward prediction error (RPE). This RPE signal, conveyed by dopamine, is thought to gate Hebbian synaptic plasticity in the striatum \citep{montague1996framework}. Over the last decade, many explicit RL models have been produced to understand the functions of dopamine and prefrontal cortex-basal ganglia circuits~\citep{cohen2009neurocomputational,maia2009reinforcement}. Recent functional magnetic resonance imaging (fMRI) studies in humans revealed that the activation in the hippocampus, a central for storing episodic memory \citep{paller2002observing}), is modulated by reward, demonstrating a link between episodic memory and RL \citep{wittmann2005reward,krebs2009novelty}. However, the existing RL models do not take into account the effect of episodic memory, which is necessary for those who want to explore decision-making by modeling circuits.

In this paper, we construct an Actor-Critic framework (\textcolor{blue}{Fig.~\ref{fig1}}, \textit{right}) based on RL theories in prefrontal cortex-basal ganglia systems (\textcolor{blue}{Fig.~\ref{fig1}}, \textit{left}) and RL algorithms for artificial systems. The Actor-Critic framework was modeled by recurrent neural network, which is a natural class of models to study mechanisms in neuroscience systems because they are both dynamical and computational \citep{mante2013context}. 
This framework was trained for two classical decision tasks, \textit{i.e.}, random dots motion (RDM) direction discrimination task \citep{roitman2002response} and value-based economic choice task \citep{padoa2006neurons}. For RDM task, a monkey is asked to arbitrarily choose the direction (left or right) of a flow of moving dots (\textcolor{blue}{Fig.~\ref{rdm_task}}a). We show that an agent reproduces qualitative results, that is, behavioral data generated by our framework can be fitted with: (i) psychometric function, a tool for analyzing the relationship between accuracy and stimulus strength (\textcolor{blue}{Fig.~\ref{rdm_task}}b, top), and (ii) chronometric function, a tool for analyzing the relationship between response time and stimulus strength (\textcolor{blue}{Fig.~\ref{rdm_task}}b, bottom). For value-based economic choice task, in which a monkey is asked to choose between two types of juice offered in different amounts (\textcolor{blue}{Fig.~\ref{fig3}}).
The activity of units in the critic network shows similar types of response observed in the orbitofrontal cortex of monkeys (\textcolor{blue}{Fig.~\ref{fig4}}). These results confirm that our framework can serve as a platform for studying diverse cognitive computations and mechanisms.

Moreover, anatomical and electrophysiological studies in animals, including humans, suggest that the episodic memory in the hippocampus is critical for adaptive behavior. Particularly, the latest research suggests that the hippocampus supports deliberation during value-based economic choice task  \citep{bakkour2019hippocampus}. Our computational framework also supports this experimental conclusion (\textcolor{blue}{Fig.~\ref{fig5}}). Yet how the brain selects experiences, from many possible options, to govern the decisions has always been an open question. To address this gap, we investigated which episodic memories should be accessed to govern future decisions by conducting experiment on this validated Actor-Critic framework in Section~\ref{investigate}. The results show that salient events sampled from episodic memories can effectively shorten deliberation time than common events in the decision-making process, suggesting that salient events stored in the hippocampus could be prioritized to propagate reward information and guide decisions.

\section{Background}\medskip
In the present work, we first trained our RNN-based Actor-Critic model using two classical decision tasks, and then conduct experiment on this optimized model to explore how episodic memory govern decision-making. The framework we designed is based on four assumptions listed below:

1. \textbf{Actor-critic architecture for RL in biological system.} This assumption states that a cortex-basal ganglia circuit (PFC-BG) can be modeled as an actor-critic architecture \citep{dayan2002reward,o2004dissociable,haber2014place}. In this process, the midbrain dopamine neurons play a central role, which code reinforcement prediction error. The actor-critic view of action selection in the brain suggests that the dorsal striatum in PFC-BG is responsible for learning stimulus-response association, which can be thought of as the `actor' in the actor-critic architecture. The ventral striatum in basal ganglia, together with cortex, is mainly used to learns state values, which is akin to the `critic'~\citep{maia2009reinforcement,maia2010two}.

2. \textbf{Recurrent neural networks reproduce neural population dynamics.} This assumption states that we can conceptualize a PFC-BG system using recurrent neural networks (RNNs), for both actor and critic. RNN is a class of artificial neural networks (ANN) with feedback connection, which has been successfully applied in both artificial intelligence~\citep{ijcai2018-98,liu2019GPN,10.1145/3390891} and computational neuroscience. There are many essential similarities between RNNs and biological neural circuits: First, RNNs units are nonlinear and numerous. Second, the units have feedback connections, which allows them to generates temporal dynamic behavior within the circuit. Third, individual units are simple, so they need to work together in a parallel and distributed manner to implement complex computations. Both dynamical and computational features of RNNs make it an ideal model for studying the mechanisms of system neuroscience \citep{rajan2016recurrent,sussillo2014neural,mante2013context}. Since basal ganglia can perform dynamic gating via reinforcement learning mechanisms (\textcolor{blue}{Fig.~\ref{fig1}}, \textit{left}), here we consider more sophisticated units, i.e., gated recurrent units (GRUs), to implement this gating mechanism. 

3. \textbf{Episodic memory contributes to decision-making process.} This assumption states that episodic memory, depending crucially on the
hippocampus and surrounding medial temporal lobe
(MTL) cortices, can be used as a complementary system for reinforcement learning to influence decisions. First, in addition to its role in remembering the past, the MTL also supports the ability to imagine specific episodes in the future \citep{hassabis2007patients}, with direct implications for decision making \citep{peters2010episodic}. Second, episodic memories are constructed in a way that allows relevant elements of a past event to guide future decisions \citep{shohamy2008integrating}.

4. \textbf{There are two different forms of learning in biological systems: slow learning and fast learning.} Many evidence suggests that cortex-basal ganglia circuits appear to implement reinforcement learning \citep{frank2004carrot}. Hence, the synaptic weights of dopamine targets (striatum in BG) in the circuit, including the PFC network, can be modulated by a model-free RL procedure. This method of incremental parameter adjustment makes it a slow form of learning. On the other hand, as mentioned above, episodic memories stored in the hippocampus impact reward-based learning, suggesting that the hippocampus can serve as a supplementary system to reinforcement learning. From this, episodic memories in replay buffer (a function similar to the hippocampus) can be used to estimate the value of actions and states to guide reward-based decision-making \citep{wimmer2014episodic}, which is a fast form of learning.

These assumptions are all based on existing research. For demonstration, we abstract the neural basis of RL in biological systems (\textcolor{blue}{Fig.~\ref{fig1}} \textit{left}) into a simple computational model (\textcolor{blue}{Fig.~\ref{fig1}} \textit{right}), an actor-critic equipped with episodic memory architecture, in which actor network leverages noisy and incomplete perceptual information about the environment to make a choice, while the critic network emits the value of the selected option. We exploit recent advances in deep RL, specifically the application of the policy gradient algorithm on RNN \citep{bakker2002reinforcement}, to train our model to perform decision-making task.

\section{Methods}\medskip
\subsection{Computational Model}\smallskip

\begin{figure*}
\centering
\includegraphics[width=0.9\textwidth]{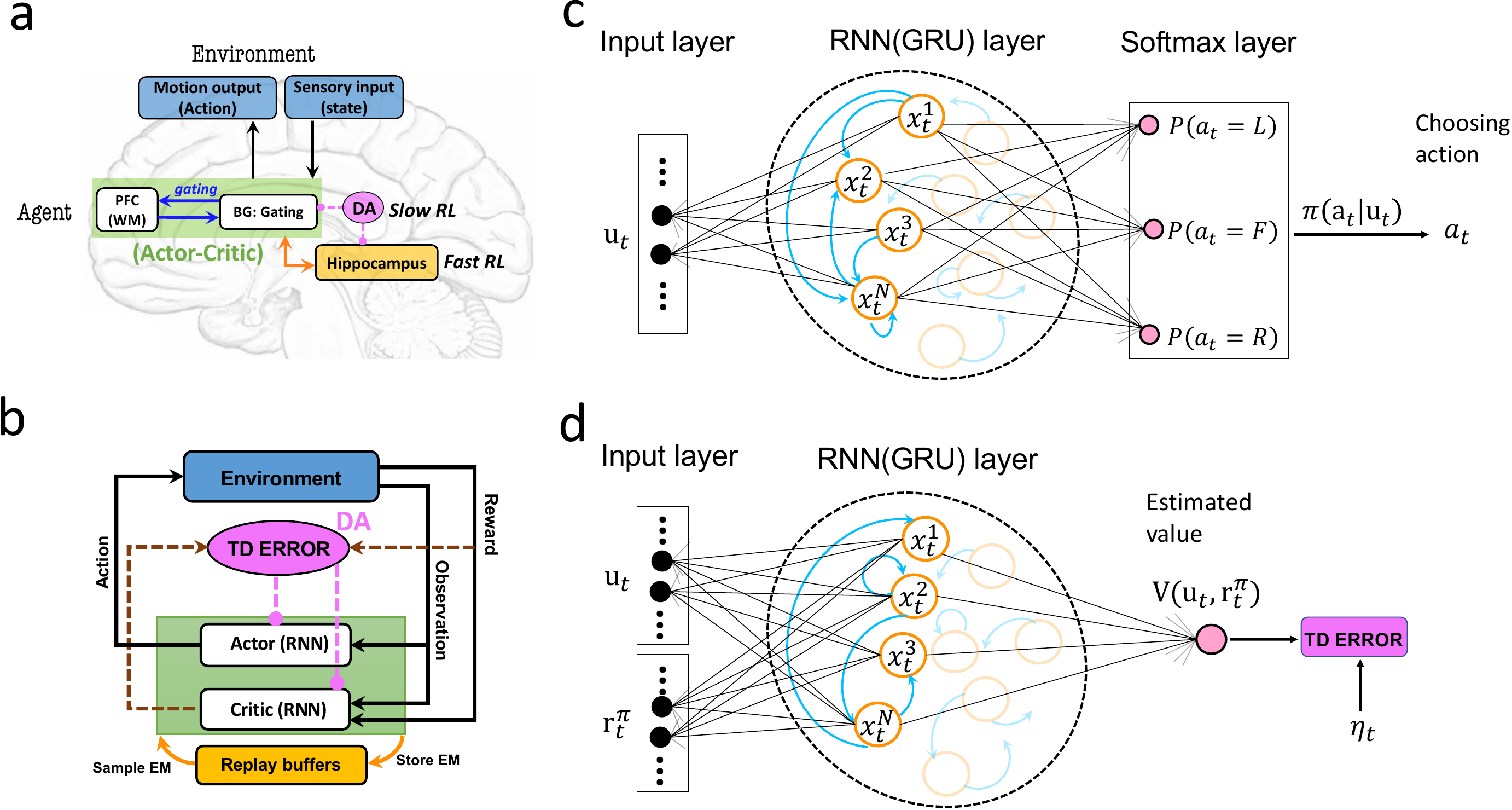}
\caption{Actor-Critic framework equipped with episodic memory. \textbf{(a)} Anatomy of a model of reinforcement learning.
The model is focused on \textbf{PFC} (robust active maintenance of task-relevant information), \textbf{BG} (dynamic gating of PFC active maintenance), \textbf{DA} (encoding a reward prediction error), \textbf{Hippocampus} (storing episodic memory). Sensory inputs are processed by PFC-BG circuit, and corresponding motor signals are sent out by Thalamus (not shown here). Working memory representations in PFC are updated via dynamic gating by the BG. These gating functions are learned by BG based on modulatory input from dopaminergic neuron (purple dotted line), \textit{i.e.}, dopamine drives reinforcement learning (slow RL) in BG regions. Moreover, dopamine modulates episodic memories in the hippocampus, supporting adaptive behaviors (fast RL). The synaptic weights in the PFC-BG network are adjusted by an RL procedure, in which DA conveys an RPE signal. \textbf{(b)} The computational model of reinforcement learning. The PFC-BG circuits in the brain were mapped to the Actor-Critic framework (green box). At each time step, the actor receives an observation from environment (corresponding to Sensory input) and selects an action (corresponding to Motion output) based on the past experience (working memory stored in RNN) and current sensory input. The reward is given followed by the chosen action and the environment moves the next state. The critic will estimate the action by computing the state-value function. Then the TD RPE (purple) is estimated through a Temporal Difference algorithm drives by DA, which adjusts the weight of the actor and critic network. Replay buffers (yellow) are used to store and replay episodic memories, similar to the function of the hippocampus. \textbf{(c)} A more detailed schematic of the actor network implementation used in our model: $\mathrm u$ represents sensory input, $\mathrm a$ represents action, and $t$ is time-step. Input units in the Actor model encode the current observation, which connect all-to-all with GRU units. The GRUs layer is composed of a fully connected set of GRU units ($N$ units shown by orange circles), which connect all-to-all with softmax layer encoding the probability of selecting each action. The critic network shown in \textbf{(d)}  has the same GRUs layer as actor network, which also receives observations as input from the environment. The output in the critic network is a linear unit encoding estimated state ${\mathrm V}$, combining with the reward ${\eta}$ to calculate TD error.} 
\label{fig1}
\end{figure*}

\textbf{RNN unit.} The Actor architecture used in our framework, which represents a particular RNN form, is depicted in \textcolor{blue}{Fig.~\ref{fig1}}c. 
RNNs have been introduced by neuroscientists into the field of neuroscience systems to describe the average firing rate of neural populations within a biological context \citep{wilson1972excitatory}. A general definition of an RNN unit is given by \cite{sussillo2014neural}:

\begin{align}
\mathrm {\tau} \frac{\mathrm d \bm{\mathrm x}}{\mathrm d \bm{t}}=-\bm {\bm{\mathrm x}}+{\bm{\mathrm W}}_{rec}\bm{\mathrm r}+{\bm{\mathrm W}}_{in}{\bm{\mathrm u}}+\bm {\mathrm b},
\label{eq:general-rnn}
\end{align}%

Where the $\bm{\mathrm x}$ is a vector, and the $i$th component is ${x}_i$, which can be viewed as the sum of the filtered synaptic currents at the soma of a biological neuron. The variable ${r}_i$ denotes the instantaneous, positive `firing rate', which is obtained by a threshold-linear activation function $[x]^{+}=max(0,x)$, the vector $\bm{\mathrm u}$ presents the external inputs provided to the network. $b_i$ is the bias each unit in the network receives, and the time constant ${\mathrm \tau}$ sets the timescale of the network. In our model, we use gated recurrent units (GRUs), a variant of the RNN architecture introduced by \cite{Chung2014Empirical}. GRUs use gating mechanisms to control and manage the flow of information between cells in the neural network. There are two main reasons for using GRUs: (1) Since the basal ganglia in the brain can perform dynamic gating via RL mechanisms, this gating mechanism can be implemented using GRUs; (2) A parallel neural system allows the biological agents to solve learning problems on a different timescale, and learning with multiple timescales have been shown to improve the performance and speed up the learning process by theoretical and modeling studies \citep{o2006making, neil2016phased}. This multiplicity of timescales is also an important feature of GRUs, as indicated by \cite{Chung2014Empirical}, in which each unit learns to adaptively capture dependencies over different time scales. In this work, we perform a little modification on the used GRUs according to Equation~(\ref{eq:general-rnn}). A continuous-time form of the modified GRUs is described as follows.

\begin{equation}
\begin{split}
\bm{\mathrm \alpha} &=\sigma(\bm{\mathrm W}^\alpha_{rec}{\bm{\mathrm r}}+\bm{\mathrm W}^\alpha_{in}{\bm{\mathrm u}}+\bm{\mathrm b}^\alpha),\\ 
\bm{\mathrm \beta} &=\sigma(\bm{\mathrm W}^\beta_{rec}{\bm{\mathrm r}}+\bm{\mathrm W}^\beta_{in}\bm{\mathrm u}+\bm{\mathrm b}^\beta),\\
{\mathrm \tau}{\frac{\mathrm d {\bm{\mathrm x}}}{\mathrm d {\bm{t}}}} &=-{\bm{ \alpha}} \circ {\bm{\mathrm x}}+{\bm{{ \alpha}}} \circ (\bm{\mathrm W}_{rec}(\bm{\beta} \circ {\bm{\mathrm r}})+\bm{\mathrm W}_{in}{\bm{\mathrm u}}+{\bm{\mathrm b}}+\sqrt{2\mathrm \tau { k_{rec}^2}}{\bm{\mathrm \xi}}),\\
{\bm{\mathrm r}}&=[{\bm{\mathrm x}}]^{+}
\end{split}
\label{equ:2}
\end{equation}

Where $\circ$ denotes the Hadamard product, $\sigma(x)=\frac{1}{1+e^{-x}}$ is the sigmoid function. The vector $\bm{\mathrm \xi}$ are independent Gaussian white noise scaled by $k_{rec}$, which present noise intrinsic to the RNN. The matrices $\bm{\mathrm W}^\alpha_{rec}$, $\bm{\mathrm W}^\alpha_{rec}$, and $\bm{\mathrm W}^\alpha_{rec}$ are $N\times N$ weight matrices of recurrent connection. While $\bm{\mathrm W}^\alpha_{in}$, $\bm{\mathrm W}^\alpha_{in}$, and $\bm{\mathrm W}^\alpha_{in}$ are $N\times N_{in}$ weight matrices of connection from input units to recurrent units. The vectors ${\bm{\mathrm b_\alpha}}$, ${\bm{\mathrm b_\beta}}$, and ${\bm{\mathrm b}}$ are bias. 

Threshold-linear activation function $[x]^{+}$ guarantees that Equation~(\ref{equ:2}) is a nonlinear dynamic system. These leaky threshold-linear units in GRUs are modulated by the time constant $\mathrm \tau$, with an update gate $\bm{\alpha}$ and reset gate $\bm{\beta}$. Based on the dynamics equation of the GRU defined above, the following section will provide a detailed description of Actor-Critic model.

\textbf{Actor-Critic model}. Based on the model constructed by \cite{Amir2019Models}, our Actor model is composed of three layers: an input layer, an RNN (GRUs) layer, and an output softmax layer. The RNN layer in our model consisted of $N=256$ GRU units, and the output layer contains three nodes (since there are $N_a=3$ actions in the RDM task and value-based choice task) (\textcolor{blue}{Fig.~\ref{fig1}}c). At each time step $t$, the input to the Actor model is current observation provided by the environment, and the outputs are the probabilities of choosing action given by the agent's policy. Here, the policy $\mathrm \pi (a_t|u_t;\theta)$ (parameterized by $\theta$) is implemented through the output of a linear readout by softmax normalization, which is determined by the activity $\mathrm r^{\pi}$ of GRU in actor network:

\begin{align}
\bm {\mathrm z}_t &=\bm {\mathrm W}_{out}^{\pi}\bm {\mathrm r}_t^\pi+\bm {\mathrm b}_{out}^{\pi},\\
\mathrm \pi(a_t=j|u_t;\theta) &=\frac{e^{(z_t)_j}}{\sum_{l=1}^{N_a}e^{(z_t)_l}},\quad         (j=1,...,N_a)
\label{equ:4}
\end{align}

Where $\bm {\mathrm W}_{out}^{\pi} \in \mathbb{R}^{N_a\times N}$ is matrix of connection weights from GRU layer to the softmax layer, $\bm{\mathrm z}_{t}$ is $N_a$ linear readouts and $\bm{\mathrm b}_{out}^{\pi}$ is $N_a$ bias. The process of action selection is carried out through random sampling from the probability distribution in equation~(\ref{equ:4}). This sampling can be considered as an abstract representation of action selection in the downstream circuitry through basal ganglia, which is the process for selecting `what to do next' in dynamic and unpredictable environments in real time.

The Critic model contains an input layer and a GRUs layer \textcolor{blue}{Fig.~\ref{fig1}}d. In particular, the inputs to the Critic model include not only the observation provided by the environment but also the activity of GRU in the actor network. The output is the state value function $\mathrm V$ (parameterized by $\theta_\mathrm v$), estimating the expected return from sensory input $\bm {\mathrm u}$ and telling the actor how good its action. The state value is predicted by the activity of GRU in Critic network through a linear readout. 

\begin{align}
\mathrm V(u_t;\theta_\mathrm v) &=\bm {\mathrm W}_{out}^{\mathrm v}\bm {\mathrm r}_t^\mathrm v+ {\mathrm b}_{out}^{\mathrm v},
\label{equ:5}
\end{align}

Where $\bm {\mathrm W}_{\mathrm out}^{\mathrm v} \in \mathbb{R}^{1\times N}$ is matrix of connection weights from GRU layer to the single linear readout layer ${\mathrm v_t}$, and ${\mathrm b}_{out}^\mathrm v$ is bias. 

The Actor network and Critic network have the same GRU structure. The GRUs layer consists of a set of interconnected GRU units (the memory part of the GRU), which is presented by $x_t^i$ in \textcolor{blue}{Fig.~\ref{fig1}}c for the ith GRU unit at time $t$. The value of each unit is updated based on the current input and the last value of all GRU units $(x_{t-1}^i, i=1,2, … ,N )$. 
In this way, GRUs layer can keep track of information about the history of past rewards and actions. In Actor model, each GRU unit takes its updated value as the current value $(x_{t-1}^i)$ and then transmits it to the softmax layer through a set of all-to-all connections. 
These connections determine the impact of each unit's output on the prediction of the next action. In Critic model, each GRU unit transmits its output to one unit (output layer of Critic model) and a scalar value is calculated, which evaluates the action value. As a whole, overall architecture will learn to perform decision-making task by learning the optimal policy using the Actor model and evaluating the action using Critic model.  

\subsection{Behavior tasks }
\begin{figure*}
\centering
\includegraphics[width=1\textwidth]{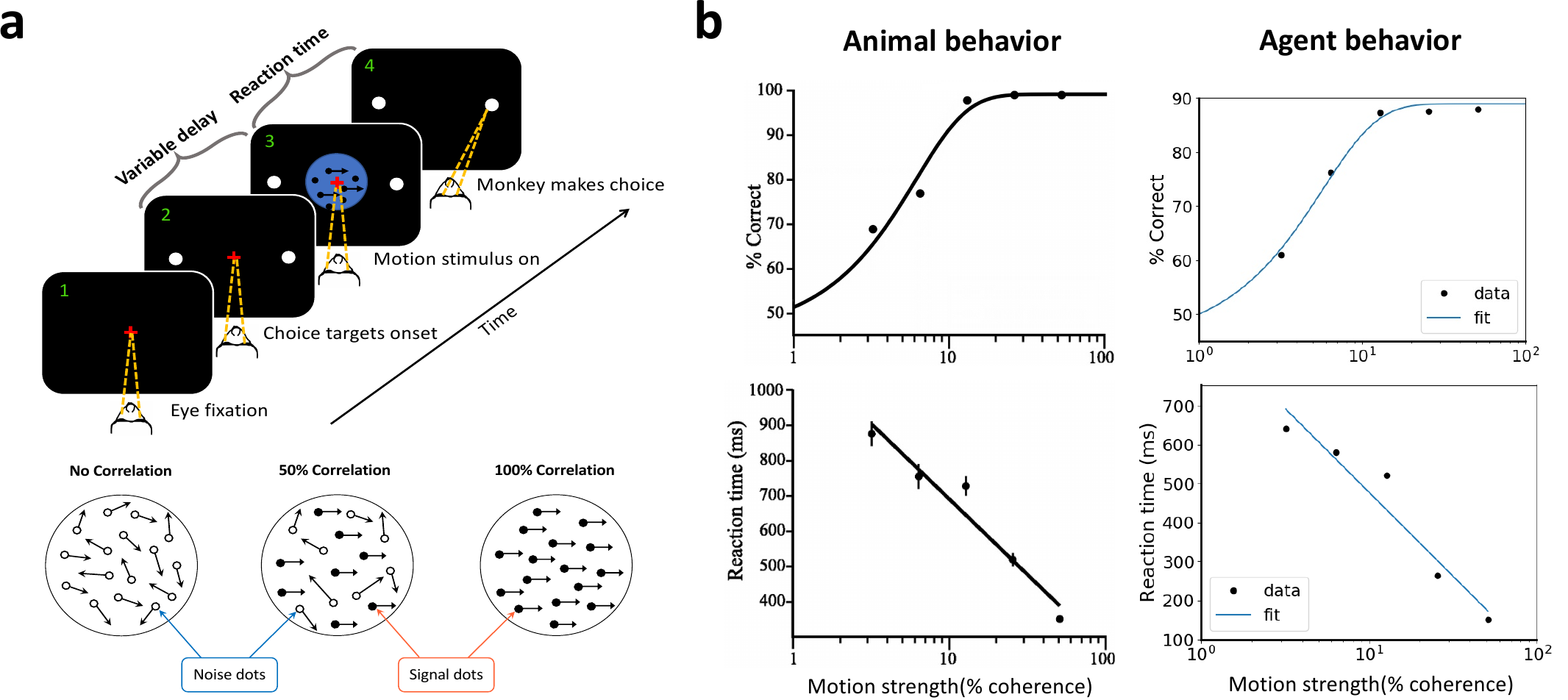}
\caption{\textbf{(a)}. RDM direction discrimination task (`reaction-time' version). Monkeys are trained to discriminate the direction of motion in a random-dot stimulus that contained coherent horizontal motion. After fixation (screen 1), the two choice targets appeared in the periphery (screen 2). After a variable delay period (was randomly selected from an exponential distribution with mean $700$ ms), dynamic random dots appeared in a $5^{\circ}$ diameter aperture (screen 3). The monkey was allowed to make a saccadic eye movement to a choice target at any time after onset of random-dot motion to indicate the direction of perceived motion (screen 4). Reaction time (RT) is defined as the elapsed time from motion onset to the initiation of the saccade, which was controlled by the monkeys and could be measured. \textit{(Buttom)} Examples of random-dot motion stimulus of variable motion coherence. Stimulus strength is varied by changing the proportion of dots moving coherently in a single direction, which determines the difficulty of the task. The lower (higher) the coherence levels, the more difficult (easier) the task is. Coherently moving dots are the `signal', and randomly moving dots are the `noise'. \textbf{(b)}. Behavior comparison of the animal and the agent. During training for the RDM task, the behaviors of the agent reflected in psychometric functions \textit{(top)} and chronometric functions \textit{(bottom)}. \textit{Left}: animal behavioral data from one experience (reproduced from \cite{roitman2002response}. \textit{Right}: our agent behavioral data. \textit{Top}: Psychometric functions from reaction time version of the RDM. The probability of a correct direction judgment is plotted as a function of motion strength and fitted by sigmoid functions. \textit{Bottom}: Effect of motion strength on reaction time (average reaction time of correct trials). The relationship between the log scaled motion strength and the reaction time fits a linear function.}
\label{rdm_task}
\end{figure*}

\smallskip
\label{sec:task}
\textbf{RDM direction discrimination task.} In the RDM discrimination task (`reaction-time' version), a monkey chooses between two visual targets; a general description is shown in~\textcolor{blue}{Fig.~\ref{rdm_task}}a. First, the monkey was required to fixate a central point until the random dot motion appears on the screen. Then, the monkey indicated its decision in the direction of dots, by making a saccadic eye movement to the target of choice. In the standard RL model, an RL agent learns by interacting with its surrounding environment and receiving rewards for performing actions. Accordingly, in the RDM task, the actual direction of the moving dots can be considered to be a state of the environment. This state is partially observable, since the monkey does not know the precise direction of the coherent motion. Therefore, the monkey needs to integrate the noisy sensory stimuli to figure out the direction. The monkey is given a positive reward, such as fruit juice, for choosing the correct target after the fixation cue turns off, while a negative reward is given, in the form of timeouts, when either the fixation is broken too early or no choice is made during the stimulus period. During the simulation, the incorrect response was rewarded with a zero reward. Given the reward schedule, the policy could be modeled and optimized using the method of policy gradient.

\begin{figure}
\centering
\includegraphics[width=0.4\textwidth]{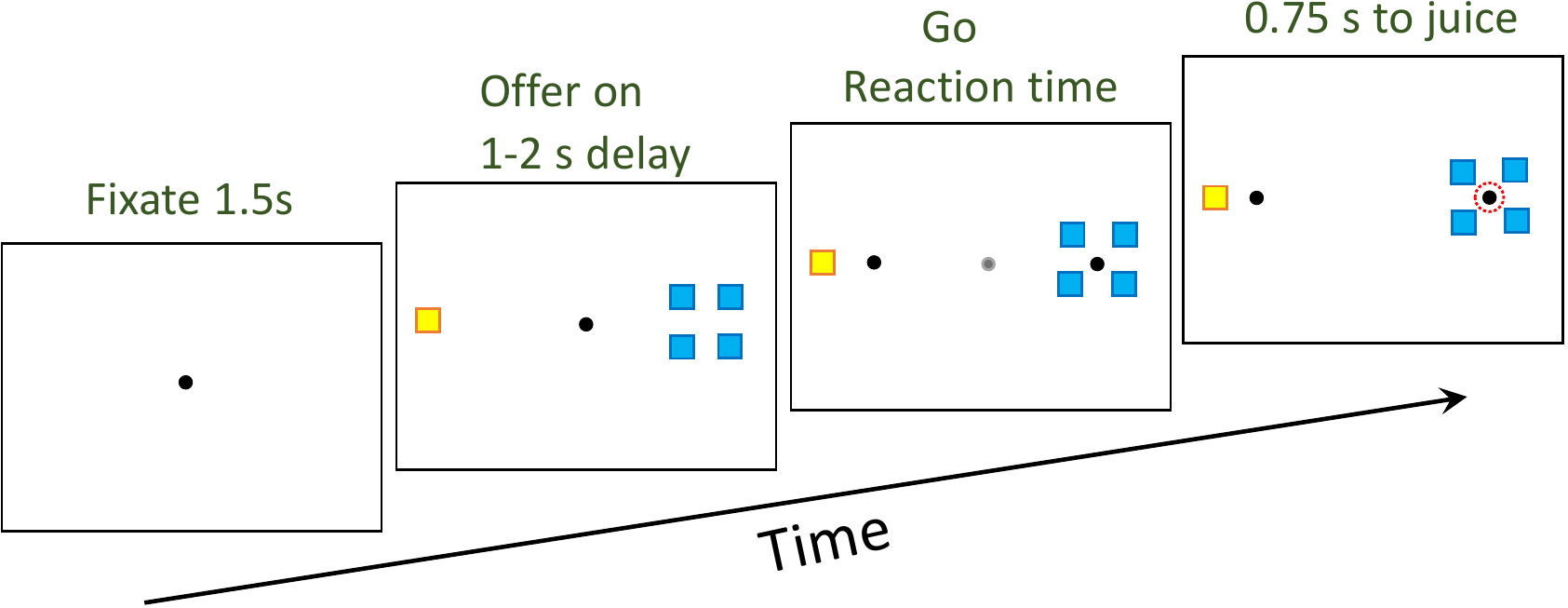}
\caption{Value-based economic choice task. At the beginning of each trial, the monkey fixated a center point on the monitor. Then two offers appeared on the two sides of the center fixation. The offers were represented by two sets of squares, with the color linked to the juice type and the number of squares indicating juice amount, which remained on the monitor for a randomly variable delay. The monkey continued fixation the center point until it was extinguished (`go' signal), at which point the monkey indicated its choice by making a saccade towards one of two targets.}
\label{fig3}
\end{figure}

\textbf{Value-based economic choice Task.} In the economic choice task experiment, reported by \cite{padoa2006neurons}, the monkey chooses between two types of juice (labeled A and B, with A being preferred) offered in different amounts \textcolor{blue}{Fig.~\ref{fig3}. Each trial began with a fixation period of $1.5s$} and then the offer, which indicated the juice type and amount for the left and right choices, was presented for $1-2s$ before it disappeared. The network was required to indicate its decision in a decision period of $0.75s$.
Since there is a choice that leads to higher rewards, in this sense, there is a `correct' answer in each trial.  

\section{Experiment}\medskip
\label{sec:5}

In this section, we will describe in detail how the Actor-Critic model learns a behavioral policy to maximize the cumulative reward. 

The interaction between a monkey and an experimentalist is regarded as the interaction between agent $\mathcal A$ and environment $\mathcal E$. At each time step $t$, the agent observes the inputs $u_t$ from the environment and then selects an action $a_t$ to be performed. The probability of selecting action $a_t$ is given by the policy function $\pi$. After performing the action $a_t$, the environment provides the agent with a scalar reward $\eta_{t}$ (here we use $\eta$ to distinguish it from ${\mathrm r}$, the firing rates of the GRU). In summary, the actor network attempts to learn a policy $\pi$ by receiving feedback from the critic network, and the critic network learns a value function $\mathrm V$ (the expected return in rewards), used to determine how advantageous it is to be in a particular state.

\subsection{Experiment 1: Training our framework to perform RDM task} \smallskip

For the RDM task, the actual direction of the moving dots can be considered to be a state of the environment. For the monkey, this state is partially observable. Learning this behavioral task by an RL algorithm is to solve a partially observable Markov decision process (POMDP). At each time $t$, an observable information is drawn from a set of environment states according to a probability distribution ${\mathrm P}(\mathrm u_t |\mathrm s_t)$. The sensory input, \textit{i.e.}, the observation received by the agent, is denoted as a tuple $\bm{\mathrm u}=(\mathrm {c_F}, \mathrm{c_L}, \mathrm{c_R})$, where $\mathrm{c_F}$ is fixation cue, $\mathrm {c_L}$ is the percentage of dots moving in the left direction, $\mathrm {c_R}$ is the percentage of dots moving in the right direction. These percentages represent the noisy evidence for two choices (left) and (right). At each time, the agent selects to perform one from the set of actions $\mathrm {A=\{{F, L, R}\}}$: fixation $(a_t=\mathrm F)$, select left $( a_t=\mathrm L)$, select right $( a_t=\mathrm R)$. A trial ends as long as the agent makes a decision (select left or right): the agent is reward with $\eta=8$ for making a correct decision and with $\eta=0$ for making a wrong decision. Aborting trial, i.e., breaking fixation early before the `go’ cue, results in a negative reward $\eta=-2$. 
If the agent has not made a choice at the maximum time $t_{max}$, the reward is $\eta=0$. Here we use $e^{-t/\tau_{\eta}}$ to discount future rewards \citep{doya2000reinforcement}, where $\tau_{\eta}$ is time constant. Discounted rewards still denote as $\eta$. Given reward function $\eta=\eta(\mathrm u_t,\mathrm a_t)$, the learning is implemented by single-threaded Advantage Actor-Critic (A2C) algorithm described by \cite{mnih2016asynchronous}. 

The goal of the agent is to learn a policy that maximizes the expected future reward to be received, starting from $t=0$ until the terminal time $T$($\leq t_{max}$). 

\begin{align}
J(\theta) &=\mathbb{E}[\sum_{t=0}^{T-1} \eta_{t+1}],
\label{equ:6}
\end{align}

For policy network, \textit{i.e.}, actor network, the loss function $\mathcal{L}^\pi (\theta)$ is defined as following. 

\begin{align}
\mathcal{L}^\pi (\theta) &=-J(\theta)+\beta_e H^{\pi}(\theta),
\label{equ:7}
\end{align}
We introduce entropy $H^{\pi}(\theta)$ to the policy network, which encourages exploration by preventing the agent from being too decisive and converging at local optima and $\beta_e$ is hyperparameter controlling the relative contribution of entropy regularization term. The key gradient $\nabla_\theta J(\theta)$ is given for each trial by the A2C algorithm.

\begin{align}
\nabla_\theta J(\theta) &=\sum_{t=0}^{T}\nabla_\theta \log \pi(\mathrm a_t|\mathrm u_t;\theta)A(u_t, r^{\pi}_t),\label{equ:8}\\
A(\mathrm u_t, \mathrm r^{\pi}_t) &=\eta_t+\gamma \mathrm V(\mathrm u_{t+1},\mathrm r^{\pi_{t+1}};\theta_ {\mathrm v})-\mathrm V(\mathrm u_{t},r^{\pi}_t;\theta_ {\mathrm v}),
\label{equ:9}
\end{align}
\noindent
where the parameters $\theta$ and $\theta_\mathrm v$ consist of connection weight, biases of the actor network and critic network respectively, \textit{i.e.}, $\theta=\{\bm {\mathrm W}_{in}^\pi,\bm {\mathrm W}_{rec}^\pi,\bm {\mathrm W}_{out}^\pi,$ $\bm {\mathrm b}_{in}^\pi,\bm {\mathrm b}_{rec}^\pi,$ $\bm {\mathrm b}_{out}^\pi\}$, $\theta_\mathrm v=\{\bm {\mathrm W}_{in}^{\mathrm v}, \bm {\mathrm W}_{rec}^{\mathrm v},\bm {\mathrm W}_{out}^{\mathrm v},\bm {\mathrm b}_{in}^{\mathrm v},\bm {\mathrm b}_{rec}^{\mathrm v},\bm {\mathrm b}_{out}^{\mathrm v}\}$. The actor learns a policy $\pi$ (the rule that the agent follows) by receiving feedback from a critic. The critic learns a state value function $\mathrm V(\mathrm u_t,\mathrm r^{\pi}_t;\theta_ {\mathrm v})$ (the expected return in rewards), which is used to determine how advantageous it is to be in a particular state by estimating the advantage function $A(\mathrm u_t,\mathrm r^{\pi}_t)$, \textit{i.e.}, TD error. The parameter $\gamma$ is the discount factor.

For value network, the loss function $\mathcal{L}^\mathrm v (\theta)$ is Mean Square Error

\begin{align}
\mathcal{L}^\mathrm v (\theta) &= \sum_{t=0}^{T}[\eta_t+\gamma \mathrm V(\mathrm u_{t+1},\mathrm r^{\pi}_{t+1})-\mathrm V(\mathrm u_{t},\mathrm r^{\pi}_t)]^2,
\label{equ:10}
\end{align}

We can get the loss function for the model overall through combining the two loss functions

\begin{align}
\mathcal{L} (\theta) &= \mathcal{L}^\pi (\theta)+\beta_\mathrm v \mathcal{L}^\mathrm v (\theta),
\label{equ:11}
\end{align}

Here, the hyperparameter $\beta_\mathrm v$ controls the relative contribution of the value estimate loss.

After every trial, the policy network and value network use Adam stochastic gradient descent (SGD) to find the parameters $\theta$ that minimizes an objective function $\mathcal{L} (\theta)$.

\begin{equation}
\begin{split}
&\nabla_\theta \mathcal{L}(\theta) = \nabla_\theta \mathcal{L}(\theta)+\beta_\mathrm v \nabla_\theta \mathcal{L}(\theta)\\
&=-\sum_{t=0}^{T}\nabla_\theta \log \pi_\theta(\mathrm a_t|\mathrm u_t)A(\mathrm u_t, \mathrm r^\pi_t)+\beta_\mathrm e \nabla_\theta H^\pi (\theta)+\beta_\mathrm v \nabla_\theta \mathcal{L}^\mathrm v (\theta),
\label{equ:12}
\end{split}
\end{equation}

The gradient $\nabla_\theta \log \pi_\theta(\mathrm a_t|\mathrm u_t)$, $\nabla_\theta H^\pi (\theta)$, $\nabla_\theta \mathcal{L}^\mathrm v (\theta)$ are computed using the backpropagation through time (BPTT). Through this training, the actor network learns to extract history experiences into the hidden state, in the form of working memory (WM). This working memory is thought to be facilitated by the PFC, which can be summarized to instruct the actor system to select rewarding actions. Meanwhile, the critic system learns a value function to train the actor network, which in turn furnishes a dynamic gating mechanism to control updating the working memory. 

\subsection{Experiment 2: Training our framework to perform value-based economic task}\smallskip

We also trained the Actor-Critic model to perform the value-based economic choice task, described in Section~\ref{sec:task}, with a training procedure similar to the above-described one for the RDM task. In this task, we noticed that there was no real correct or wrong choice for the monkey. However, there is a choice that allowed the monkey to receive the highest reward, this choice can thus be considered as a 'correct' choice. Unlike the RDM task, the information regarding whether an answer is correct is not included in the inputs, but rather in the correlation between the inputs and rewards.

\subsection{Test behavioral characteristics of our framework}\smallskip
\label{sec:test}

Next, we investigated whether the Actor-Critic framework captures the behavioral characteristics of animals in the cognitive experiments. In the previous section, we have trained the Actor-Critic framework to perform the RDM and value-based economic choice tasks. Here, we compare the behavioral characteristics exhibited by the trained model with those observed in the animal experiments.\smallskip

\textbf{RDM task}. The results are consistent with the behavioral findings from the animal experiments, which are mainly reflected in the psychometric and chronometric functions, as shown in \textcolor{blue}{Fig.~\ref{rdm_task}}b.
The performance accuracy in the RDM task depends on the strength of the sensory input, and the psychometric function is a good tool to analyze such a relationship. The percentage of correct direction judgments is plotted as a function of the motion strength (measured by the proportion of coherently moving dots). \textcolor{blue}{Fig.~\ref{rdm_task}}b \textit{(top)} shows a high accuracy during a strong motion, while less accuracy is shown with more chance and a weaker motion, which suggests that the agent in our Actor-Critic framework captures this important behavioral feature. Moreover, the theory of chronometric functions puts a constraint on the relationship between the response time and accuracy. A difficult task (weaker stimuli strength) requires the agent to take more time to make a decision (\textcolor{blue}{Fig.~\ref{rdm_task}}b \textit{(bottom)}), which means that the additional viewing time for difficult trials was devoted to integrating the sensory information. As a result, the appropriate trade-off between speed and accuracy is learned by this Actor-Critic framework. It is worth emphasizing that unlike the usual machine learning goals, our objective is not to achieve the 'perfect' performance, but rather to train the agents to match the smooth psychometric and chronometric characteristics observed in the behavior of the monkeys.\smallskip

\begin{figure*}
\centering
\includegraphics[width=1\textwidth]{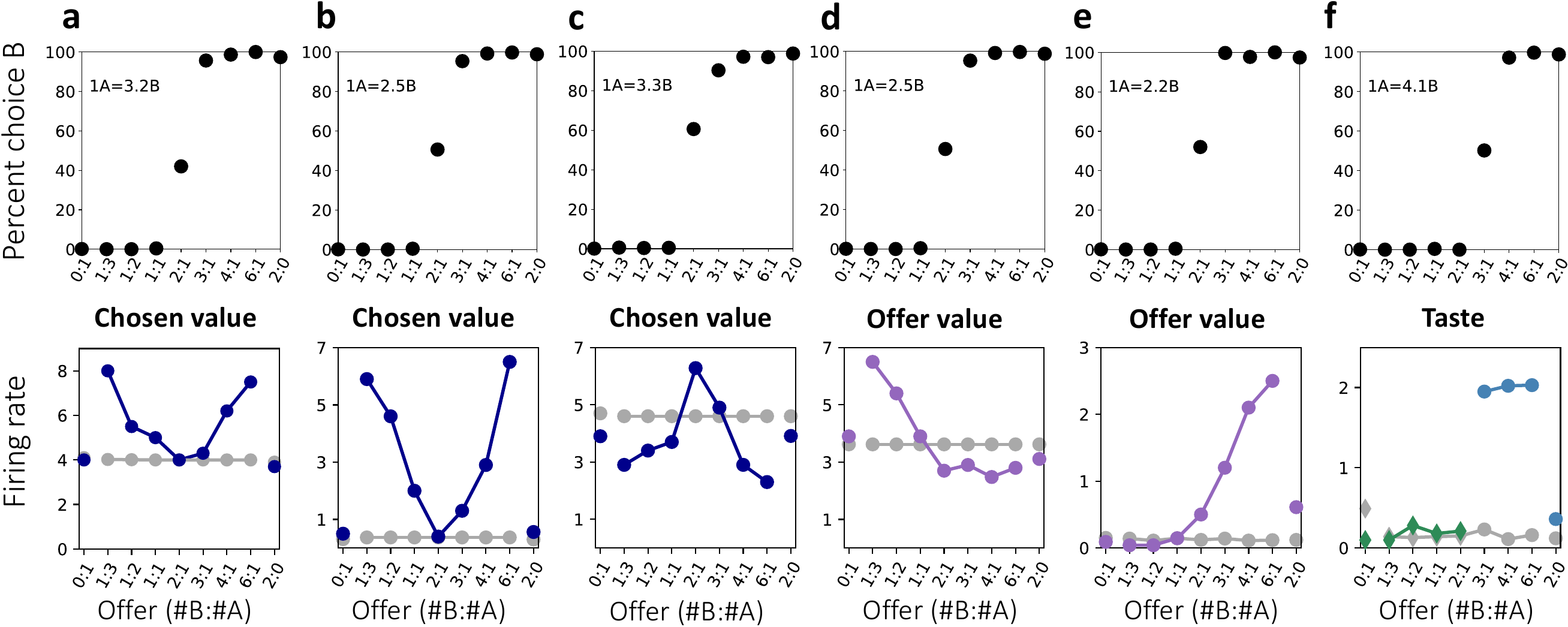}
\caption{The units in our model exhibit diverse selectivity for the task variables,  as observed in the orbitofrontal cortex. The top panel shows the percentage of trials in which the agent chose `juice' B ($y$ axis) for various offer types ($x$ axis). The relative value is indicated on the top left. For example, in \textbf{a}, the relative value is $3.2$, which indicates that the reward contingencies are indifferent between $1$ `juice' of A and $3.2$ `juice' of B. Different relative values indicate different choice patterns. The bottom panel of the figure shows the mean activity ($y$ axis, defined as the period of $800$ ms before the decision) of example value network units for various offer types ($x$ axis) under different choice patterns: $1$A = $3.2$B (\textbf{a}, deep blue), $1$A = $2.5$B (\textbf{b}, deep blue), $1$A = $3.3$B (\textbf{c}, deep blue), $1$A = $2.5$B (\textbf{d}, purple), $1$A = $2.2$B (\textbf{e}, purple), and $1$A = $4.1$B (\textbf{f}, green and blue). For each case, the grey circles show the mean the activity of value network units during the fixation period. \textbf{a-c}, The units in the value network exhibit selectivity for the `chosen value'. \textbf{d-e}, The units in the value network exhibit selectivity for the `offer value'. \textbf{f}, The trials are separated into choice A (green diamonds) and choice B (blue circles). 
}
\label{fig4}
\end{figure*}

\textbf{Value-based economic choice task}. The activity of the units in the critic network exhibits similar types of response to those observed in the orbitofrontal cortex of the monkeys \citep{padoa2006neurons}. First, roughly $20\%$, $60\%$, and $20\%$ of the active units are selective to the chosen value, the offered value, and to choose alone, respectively, as defined in the animal experiment. Second, there is a trade-off between the juice type and its quantity (upper panel of \textcolor{blue}{Fig.~\ref{fig4}}). Third, the patterns of neural activity are consistent with the behavioral findings from the animal experiment, with three main response patterns: (i) similar U-shaped response pattern (\textcolor{blue}{Fig.~\ref{fig4}}a-c, {deep blue circles}); (ii) the response pattern associated with the `offer value' variable (\textcolor{blue}{Fig.~\ref{fig4}}d-e, purple circles); (iii) the response pattern related to the juice `taste' variable. For this task, the network architecture has not been changed, and we only change the initial value of the critic network's input weight.

\begin{table}[width=.9\linewidth,cols=4,pos=h]
\caption{Parameter for Actor-Critic model training.}\label{tbl1}
\begin{tabular*}{\tblwidth}{@{} LLLL@{} }
\toprule
Parameter & Value & Parameter & Value\\
\midrule
Learnig rate & 0.004 & $t_{max}$ & 275 \\
$\tau$ & 50ms & $k_{rec}^2$ & 0.01 \\
$\tau_{\eta}$ & 200ms & $\beta_\mathrm v$ & 0.5 \\
$\gamma$ & 0.99 & $\beta_\mathrm e$ & 0.5 \\
\bottomrule
\label{table1}
\end{tabular*}
\end{table}

\section{Analysis}\medskip
In Section~\ref{sec:test}, which suggests that it can serve as a computational platform to study the impact of memory on the cognitive function. It has been shown by a number of experimental studies that memory is essential to make decisions, enabling the organisms to predict possible future outcomes by drawing on past events. For instance, working memory, which is a temporary storage in the brain \citep{repovvs2006multi}, has been shown to guide the choice by maintaining and manipulating task-relevant information. Besides, episodic memory has also been shown to be involved in the decision-making process. Moreover, a recent study suggests that the hippocampus supports deliberation about the value during the value-based economic choice task: thus, the hippocampus contributes to the construction of internal samples of evidence that are related to decision-making \citep{bakkour2019hippocampus}. Based on this idea, in this section, we combine our computational platform with the value-based economic choice task to explore the role of episodic memory in the process of decision-making.

\subsection{Episodic memory contributes to decision-making}\smallskip
First, we need to verify whether the Actor-Critic model that is equipped with episodic memory has an effective performance. Psychologically, episodic memory refers to the capacity to consciously recollect an autobiographical memory of the events that occurred in particular times and places. For example, a person can recall an episode from the past, such as his $20^{\rm th}$ birthday party, and remember who was there and where it happened. Computationally, we mainly emphasize the notion of one-time episodes (like one-trial learning in a task). A previous study suggested that episodic memory could be used to store the specific rewarding sequence of state-action pairs and later try to mimic such a sequence, a process called episodic control \citep{lengyel2008hippocampal}. In this work, we propose a slightly different computational principle, in which episodic memory is used to optimize the policy rather than directly extract it.

In our computational model, one episodic memory is generated as follows: On each trial $i$ in the value-based economic choice task, the agent's experiences $e_t=(\mathrm u_t,\mathrm a_t,\eta_t,\mathrm s_{t+1})$ at each time step $t$ are stored as an episodic memory $E_i=(\mathrm u_0,\mathrm a_0,\eta_0,\mathrm s_1,…,\mathrm u_t,$ $\mathrm a_t,\eta_t,\mathrm s_{t+1},…,\mathrm u_{T_{i-1}},\mathrm a_{T_{i-1}},\eta_{T_{i-1}},$ $\mathrm s_{T_{i}})$ and $T_i$ is the length of the $i$th trial. According to the reward received at the end of the $i$th trial, we can divide the memory into three types: the trial with positive reward (denoted as $E_i^{posi}$), the trial with negative reward~(denoted as $E_i^{nega}$), and the trial with zero reward (denoted as $E_i^{zero}$). Then the agent stores these episodic memories in one replay buffer $D=\{\{E_1^{posi},…,E_{N_1}^{posi}\},$ $\{E_1^{nega},…,E_{N_2}^{nega}\},$ $\{E_1^{zero},…,E_{N_3}^{zero}\}\}$, a pool of memories, the function of which is similar to the hippocampus in the brain. 

How does past experience stored in replay buffer optimize behavior policy? At the computational level, a method called importance sampling can be used to estimate the expected return $J(\theta)$ by sampling episodic memory from replay buffer $D$. In fact, this behavior policy for collecting samples is a known policy (predefined just like a hyperparameter), labeled as $\mu(\mathrm a|\mathrm u)$. Suppose we retrieve a single experience $(\mathrm u_0,\mathrm a_0,\eta_0,\mu(.|\mathrm u),…,\mathrm u_t,\mathrm a_t,\eta_t,\mu(.|u_t)$ $,…,$ $\mathrm u_{T_{i-1}}$,
$\mathrm a_{T_{i-1}},\eta_{T_{i-1}},\mu(.|\mathrm u_{T_{i-1}}))$, where actions have been sampled from episodic memory according to the behavior policy $\mu(a|u)$. Given that the training observations, the policy gradient can be rewritten as:

\begin{align}
\nabla_\theta J(\theta) &=\sum_{t=0}^{T}\frac{\pi ( \mathrm a_t|\mathrm u_t;\theta)}{\mu(\mathrm a_t|\mathrm u_t)}\nabla_\theta \log \pi(\mathrm a_t|\mathrm u_t;\theta)A(\mathrm u_t, \mathrm r^\pi_t),
\label{equ:13}
\end{align}
\noindent where $\frac{\pi(\mathrm a_t|\mathrm u_t;\theta)}{\mu(\mathrm a_t|\mathrm u_t)}$ is the importance weight, and $\mu$ is non-zero whereever $\pi(\mathrm a_t|\mathrm u_t;\theta)$ is.  
We note that in the case where $\frac{\pi_\theta (\mathrm a_t|\mathrm u_t;\theta)}{\mu(\mathrm a_t|\mathrm u_t)}=1$ the equation~(\ref{equ:13}) is the same as equation~(\ref{equ:8}). To use episodic memory to optimize policy, we define the learning process as follows: for trial $n=1$, policy network was updated with equation~(\ref{equ:12}), in which the gradient term $\nabla_\theta J(\theta)$ is represented by equation~(\ref{equ:8}). Then the agent store full trajectory (an episodic memory) of this trial in replay buffer. For the trial $n=2$, the agent randomly samples a trajectory as past experience to optimize policy and the gradient term $\nabla_\theta J(\theta)$ is represented by equation~(\ref{equ:13}). These steps are repeated until the training terminal, at which point the agent learns a policy concerning how to perform the value-based economic choice task.

\begin{figure}
\centering
\includegraphics[width=0.45\textwidth]{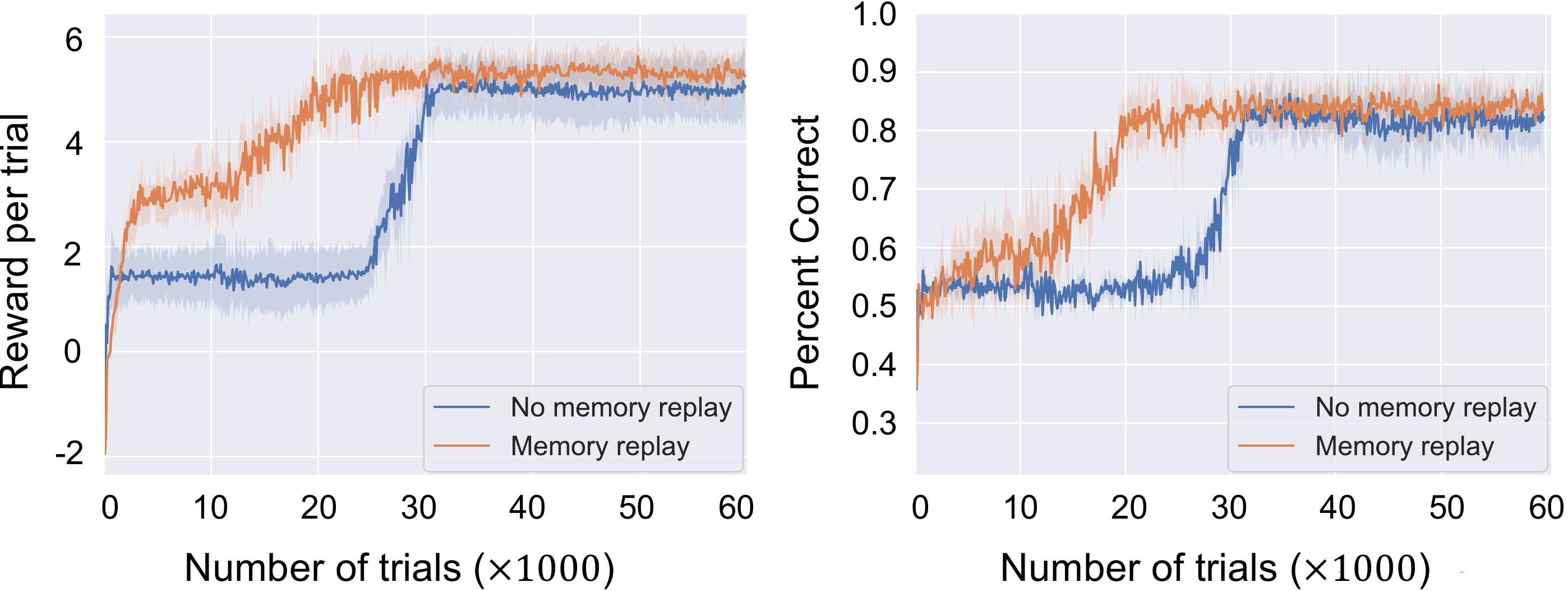}
\caption{Learning curves of the agent with episodic memory (orange line) and without episodic memory (blue line) on the economic choice task. \textit{(Left)} Average reward per trial. \textit{(Right)} Percent correct, for trials on which the network made a decision.
}
\label{fig5}
\end{figure}

\textcolor{blue}{Fig.~\ref{fig5}} \textit{(left)} shows the learning curve of agents with and without episodic memory (orange line and blue line, respectively) for the value-based economic choice task (the average return of $2000$ trial samples). It can be seen that the agent with episodic memory performs significantly faster in this task compared with the one without episodic memory, although both policies eventually reached the same performance. These results are consistent with some recent studies showing that animal decisions can indeed be guided by samples of the individual past experience ~\citep{murty2016episodic}.

The percentage of correct trials is shown in \textcolor{blue}{Fig.~\ref{fig5}} \textit{(right)} and it is calculated by $N_{right}/N_{choice}$, where $N_{choice}$ represents the number of trials in which the monkey made a choice (right or error) in $20000$ trials, and $N_{right}$ denotes the number of correct choices. It can be observed that at the beginning of the trial, the correct percentage of agents who cannot extract episodic memory from the replay buffer is maintained at around $50\%$ (blue line), and only after substantial training (about $30000$ trials) can the agent achieve the baseline accuracy rate. This suggests that the agent equipped with episodic memory shows a better execution efficiency.

\subsection{Episodic memory for salient event}\smallskip
\label{investigate}

In the previous section, we have verified that episodic memory indeed allows the agent to learn a task faster. Nevertheless, the question of which types of episodic memory samples should be selected to govern the decisions remains unanswered in the field of cognitive neuroscience. In this section, we will examine this question.

The relationship between events is often clear only when they are reviewed. For example, when something positive happens, we want to know how to repeat this event. However, when an event occurs before the reward is given, how to know what causes it? This is the earlier mentioned `temporal credit assignment problem', which can be solved by saving all the potential determinants, such as rewards, of behaviorally relevant events into working memory. We propose the question of how does episodic memory balance the need to represent these potential determinants of reward outcomes to deal with credit assignment? One solution may be to enhance episodic memory for notable events, referred to as 'salient memory', which are potential reward determinants. In fact, both the violations and conformance of expectancy can be considered as salient events to be stored in the memory buffer. Since such long-term memories are potentially predictive of reward outcomes, it will provide a computationally feasible way to obtain future rewards.

In the value-based economic choice task, salient events include trials in which the right choice was made (rewarded; expectancy conformance) or the fixation was broken (punished; expectancy violation). When it comes to a gaze-breaking trial, the agent's policy cannot be optimized due to insufficient interaction with the environment. As a result, we only choose expectancy conformance as a salient event. In the third type of trials, the monkeys made a response before the trial was over, but their choice was wrong. The incorrect response was neither rewarded by the juice nor punished. Such a trial can be considered as a common event, because it's not a particular event for monkeys. Accordingly, the episodic memories in the replay buffer $D$ have three types: the set $D_{posi}=\{E_1^{posi},$ $…,E_{N_c}^{posi}\}$ for salient events, the set $D_{zero}=\{E_1^{zero},…,E_{N_z}^{zero}\}$ for common events and the remaining events are denoted as the set $D_{nega}=\{E_1^{nega},…,$ $E_{N_e}^{nega}\}$.

\begin{figure}
\centering
\includegraphics[width=0.45\textwidth]{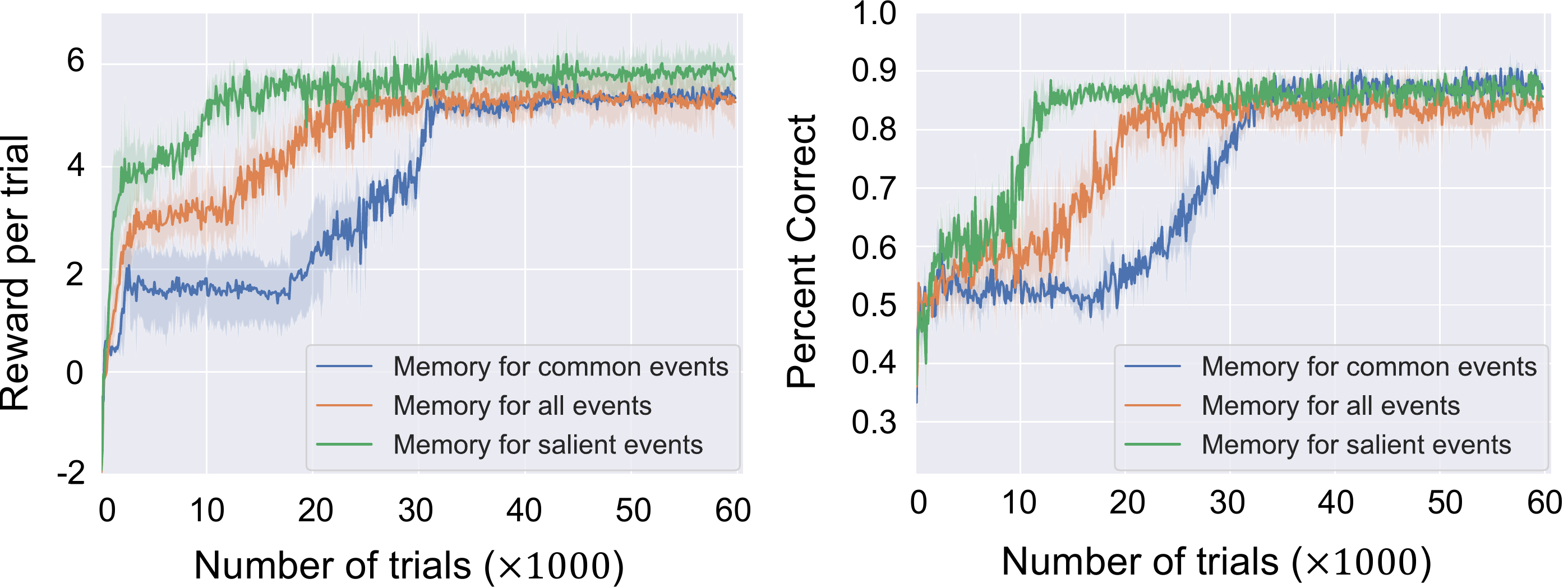}
\caption{Learning curves of an agent on the RDM task for different types of episodic memory, salient memory (green line), common episodic memory (blue line), all type of episodic memory (orange). \textit{(Left)} Average reward per trial. \textit{(Right)} Percent correct.
}
\label{fig6}
\end{figure}

In order to investigate if salient events sampled from the memory buffer can more effectively have a bias towards reward-guided choice compared with common events, we plot the learning curve of the agent with different types of episodic memories. By comparing the green (salient events) and blue (common events) curves in \textcolor{blue}{Fig.~\ref{fig6}}, we can see that the agent with significant events achieves a better performance than the agent with common events.
As shown in \textcolor{blue}{Fig.~\ref{fig6}} (\textit{left}), when the agent uniformly at random draws an event from the set $D_{posi}$ to optimize the policy, the return received by the agent can reach the baseline level more quickly (green line). However, when the agent extracts common events from the set $D_{zero}$ (blue line), it must go through a long period of learning to get higher returns. In this case, the percentage of agents who chose correctly is also always maintained at around $50\%$ at the beginning of the experiment (green line in \textcolor{blue}{Fig.~\ref{fig6}} (\textit{right})), which indicates that the monkey chooses the direction at random. As the training increases, the monkey makes more and more correct choices. It can be noted that its learning curve is similar to that of an agent who does not use memory to optimize their strategies (blue line in \textcolor{blue}{Fig.~\ref{fig5}}).
This suggests that episodic memory about common events did not help the monkeys to make choices. Moreover, when an experience is sampled from the set $D$, the reward value and final accuracy obtained by the agent are higher than those in the case where experience is sampled from the set $D_{zero}$, but lower than the case where experience is sampled from the set $D_{posi}$. Although the learning time significantly varies, the agent ends up with the same return value and accuracy in all the cases. Our results suggest that memory encoding may be stronger for trials that involved salient events. That is, the salient episodic memory in the hippocampus is more likely to be sampled during the ensuing choice.


\section{Discussion}\medskip

The goal of the present work was twofold: First, we trained an Actor-Critic RL model to solve tasks that are analogous to the monkey's tasks. This can reproduce the main features of the behavioral data so that we conduct other behavioral experiments in this framework. Specifically, we used RNNs to construct the Actor-Critic RL framework based on RL theories of the PFC-BG circuit. The model was evaluated in two classical decision-making tasks --- a simple conceptual decision-making task and a value-based economic choice task --- and successfully reproduced the behavioral features reported by \citep{shadlen2001neural} and neural activity recorded from the animal brain reported by \citep{padoa2006neurons}. We presented a computational platform, in which corresponding circuit mechanisms can be studied by systematically analyzing a model network. In addition, diverse cognitive functions can also be explored by conducting corresponding behavioral experiments. Second, based on our modeling work, we investigated which experiences in the hippocampus are ultimately considered or ignored during deliberation to govern future choices.

Since 1995, numerous actor-critic models for reinforcement learning have been proposed in the field of neuroscience, particularly in the rat’s basal ganglia \citep{davis1995models,joel2002actor}. Some evidence shows that neurons in the PFC \citep{fujii2005time} and striatum \citep{barnes2005activity} code the action sequences, suggesting that the BG-PFC circuit may participate in abstract action representations. Therefore, at the biological analysis level, our model supports the actor-critic picture for reward-based learning in the PFC-BG circuit: One circuit learns an action selection policy and implement it, while the second structure computes the expected return and offers immediate feedback that tells it whether the current action is good or bad. Moreover, \cite{frank2006anatomy} have demonstrated that the BG can implement an adaptive gating mechanism, which allows task-relevant information to be maintained into working memory (a temporary storage in the brain, and facilitated by the prefrontal cortex). Our model also supports this division of labor between PFC and BG as follows: The actor network learns task-relevant information and saves it into the hidden state in the form of working memory, while the critic system learns a value function to train the actor network, which in turn furnishes a dynamic gating mechanism to control updating the working memory.

Moreover, a recent experimental work in humans has shown that during memory-based decision-making tasks, the medial frontal cortical neurons phase-locked their activity to theta frequency band oscillations in the hippocampus, which suggests an oscillation-mediated activity coordination between distant brain regions \citep{Minxha2020Flexible}. This functional interaction between the frontal cortical and hippocampus supports our computational framework: The Actor-Critic model uses working memory stored in the hidden state of the GRU to make a choice, and this selected action affects the storage of memories in the hippocampus, which is in turn used to optimize the policy and control working memory updates. Although we have used the GRU to model the decision and value networks, both the ability of dynamic gating mechanism and storing states as working memory make our model shows a powerful computational learning performance. However, early work demonstrated that the capacity of working memory is limited, which results in decisions that are often made with finite information. Due to the transient characteristic caused by the capacity limitation and fast decay rate of working memory, it is not an ideal memory system to independently support decision-making. Moreover, accumulating evidence indicates that dopamine can facilitate episodic memory in the hippocampus encoding to support adaptive behavior \citep{bethus2010dopamine}, which suggests that episodic sampling is may be a powerful decision-making mechanism. Therefore, we investigated the link between episodic memory and reward-based choice in our framework by conducting the value-based economic choice task in our framework. The results suggest that a retrieval of salient episodic memory can promote deliberation in the decision-making process, which is essential to future goal-directed behavior.

Our model has some limitations, which may be opportunities for future work. For instance, during the retrieval of samples from episodic memories, we have defined the priority of salient events only in an abstract way, while we have not provided a mechanism to explain how the mammalian brain would compute it. Therefore, there is a need to develop a process-level model to implement this term. Moreover, in the cerebral cortex of mammals, one neuron releases only a single transmitter, known as `Dale's Principle', which generates the same effect (excitatory or inhibitory) at all of its synaptic connections to other cells. In our framework, due to the complex nature of the GRU, we omitted such a biological constraint and instead used the firing rate units as a mixture of excitatory and inhibitory neurons. In future work, it is required to reintroduce these constraints, and other physiologically relevant phenomena, such as bursting, adaptation and oscillations, may also be incorporated to build a more biologically-plausible model.

\noindent
\textbf{Acknowledgement} This work was supported by the National Natural Science Foundation of China (Grant No.11572127 and No.11172103).

\printcredits

\bibliographystyle{cas-model2-names}

\bibliography{ms}


\end{document}